\documentclass[twocolumn,showpacs,preprintnumbers,amsmath,amssymb]{revtex4}

\usepackage{graphicx}
\usepackage{dcolumn}
\usepackage{bm}

\begin{document}


\title{Characterization of maximally entangled two-qubit states
via the Bell-Clauser-Horne-Shimony-Holt inequality}
\author{Zeqian Chen}
\email{zqchen@wipm.ac.cn}
\affiliation{%
Wuhan Institute of Physics and Mathematics, Chinese Academy of
Sciences, P.O.Box 71010, 30 West District, Xiao-Hong Mountain,
Wuhan 430071, China}%

\date{\today}

\begin{abstract}
Maximally entangled states should maximally violate the Bell
inequality. In this paper, it is proved that all two-qubit states
that maximally violate the Bell-Clauser-Horne-Shimony-Holt
inequality are exactly Bell states and the states obtained from
them by local unitary transformations. The proof is obtained by
using the certain algebraic properties that Pauli's matrices
satisfy. The argument is extended to the three-qubit system. Since
all states obtained by local unitary transformations of a
maximally entangled state are equally valid entangled states, we
thus give the characterizations of maximally entangled states in
both the two-qubit and three-qubit systems in terms of the Bell
inequality.
\end{abstract}

\maketitle

The Bell inequality [1] was originally designed to rule out
various kinds of local hidden variable theories. Precisely, the
Bell inequality indicates that certain statistical correlations
predicted by quantum mechanics for measurements on two-qubit
ensembles cannot be understood within a realistic picture based on
Einstein, Podolsky, and Rosen's (EPR's) notion of local realism
[2]. However, this inequality also provides a test to distinguish
entangled from nonentangled quantum states. In fact, Gisin's
theorem [3] asserts that all entangled two-qubit states violate
the Bell-Clauser-Horne-Shimony-Holt (Bell-CHSH) inequality [4] for
some choice of spin observables.

As is well known, maximally entangled states, such as Bell states
and GHZ states [5], have become a key concept in the nowadays
quantum mechanics. On the other hand, from a practical point of
view maximally entangled states have found numerous applications
in quantum information [6]. A natural question is then how to
characterize maximally entangled states. There are extensive
earlier works on maximally entangled states [7], however, this
problem is far from being completely understood today. It is well
known that maximally entangled states should maximally violate the
Bell inequality [8]. Therefore, for characterizing maximally
entangled states, it is suitable to study the states that
maximally violate the Bell inequality. In the two-qubit case, Kar
[9] has described all states that maximally violate the Bell-CHSH
inequality. Kar mainly made use of an elegant technique, which was
originally introduced in [10], based on the determination of the
eigenvectors and eigenvalues of the associated Bell operator.

In this paper, it is proved that Bell states and the states
obtained from them by local unitary transformations are the unique
states that violate maximally the Bell-CHSH inequality. The
techniques involved here are based on the determination of local
spin observables of the associated Bell operator. We show that a
Bell operator presents a maximal violation on a state if and only
if the associated local spin observables satisfy the certain
algebraic identities that Pauli's matrices satisfy. Consequently,
we can find those states that show maximal violation, which are
just the states obtained from Bell states by local unitary
transformations.

The method involved here is simpler (and more powerful) than one
used in [9] and can be directly extended to the $n$-qubit case. We
illustrate the three-qubit case and show that all states that
violate the Bell-Klyshko inequality [11] are exactly GHZ states
and the states obtained from them by local unitary
transformations. This was conjectured by Gisin and
Bechmann-Pasquinucci [8]. Since all states obtained by local
unitary transformations of a maximally entangled state are equally
valid entangled states [12], we thus give the characterizations of
maximally entangled states in the two-qubit and three-qubit
systems via the Bell-CHSH and Bell-Klyshko inequalities,
respectively.

Let us consider a system of two qubits labelled by $1$ and $2.$
Let $A,A'$ denote spin observables on the first qubit, and $B,B'$
on the second. For $A^{(\prime)}= \vec{a}^{(\prime)} \cdot
\vec{\sigma}_1$ and $B^{(\prime)} = \vec{b}^{(\prime)} \cdot
\vec{\sigma}_2,$ we write$$(A,A') = (\vec{a}, \vec{a}' ), A \times
A' = ( \vec{a} \times \vec{a}') \cdot \vec{\sigma}_1,$$and
similarly, $(B,B')$ and $B \times B'.$ Here $\vec{\sigma}_1$ and
$\vec{\sigma}_2$ are the Pauli matrices for qubits $1$ and $2,$
respectively; the norms of real vectors $\vec{a}^{(\prime)},
\vec{b}^{(\prime)}$ in ${\bf R}^3$ are equal to $1.$ We write
$AB,$ etc., as shorthand for $A \otimes B$ and $A = AI_2,$ where
$I_2$ is the identity on qubit 2.

Recall that the Bell-CHSH inequality is
that\begin{equation}\langle A B + A B' + A' B - A' B' \rangle \leq
2,\end{equation}which holds true when assuming EPR's local realism
[2]. We define the two-qubit Bell operator
[10]\begin{equation}{\cal B}_2 = A B + A B' + A' B - A'
B'.\end{equation}Since$$AA' = (A,A') + i A \times A', A'A = (A,
A') - i A \times A',$$$$BB'= (B,B') + i B \times B', B'B= (B,B') -
i B \times B',$$a simple computation yields that
\begin{equation}{\cal B}^2_2 =4 - [A,A'][B,B']= 4 + 4 (A \times A')
(B \times B').\end{equation}Since\begin{equation}\|A \times
A'\|^2=1-(A,A')^2, \|B \times B'\|^2=1-(B,B')^2,\end{equation}it
concludes that ${\cal B}^2_2 \leq 8$ and, $\| {\cal B}^2_2 \| = 8$
if and only if\begin{equation}(A,A')= (B,B')=0.\end{equation}
Accordingly, the Bell-CHSH inequality can be violated by quantum
states by a maximal factor of $\sqrt{2}$ [13]. In particular, one
concludes that Eq.(5) is a necessary and sufficient condition that
there exists a two-qubit state that maximally violates the
Bell-CHSH inequality, i.e.,\begin{equation}\langle \psi | {\cal
B}_2 | \psi \rangle = 2 \sqrt{2}\end{equation}for some state
$\psi.$

As follows, we show that every state $| \psi \rangle$ satisfying
(6) can be obtained by a local unitary transformation of the Bell
states. Indeed, let $A''= A \times A'$ and $B''=B \times B'.$
Since Eq.(5) holds true, it concludes that both
$(\vec{a},\vec{a}',\vec{a}'')$ and $(\vec{b},\vec{b}',\vec{b}'')$
are triads in $S^2,$ the unit sphere in ${\bf R}^3.$ Then, it is
easy to check that\begin{equation}AA'=-A'A=iA'',\end{equation}
\begin{equation}A'A''=-A''A'=iA,\end{equation}
\begin{equation}A''A=-AA''=iA';\end{equation}
\begin{equation}A^2=(A')^2=(A'')^2 = 1.\end{equation}
Hence, $\{A,A',A''\}$ satisfy the algebraic identities that
Pauli's matrices satisfy [14] and similarly, $\{B,B',B''\}.$
Therefore, choosing $A''$-representation $\{|0 \rangle_A,|1
\rangle_A \},$ i.e.,
\begin{equation}A'' |0 \rangle_A = |0 \rangle_A, A''|1 \rangle_A =
-|1 \rangle_A,\end{equation}we have that\begin{equation}A |0
\rangle_A =e^{-i \alpha} |1 \rangle_A, A |1 \rangle_A = e^{i
\alpha} |0 \rangle_A,\end{equation}
\begin{equation}A' |0
\rangle_A = ie^{-i \alpha} |1 \rangle_A, A' |1 \rangle_A =-i e^{i
\alpha} |0 \rangle_A,\end{equation}$(0\leq \alpha \leq 2 \pi).$
Similarly, we have that
\begin{equation}B |0
\rangle_B =e^{-i \beta} |1 \rangle_B, B |1 \rangle_B = e^{i \beta}
|0 \rangle_B,\end{equation}
\begin{equation}B' |0
\rangle_B = ie^{-i \beta} |1 \rangle_B, B' |1 \rangle_B =-i e^{i
\beta} |0 \rangle_B,\end{equation}
\begin{equation}B'' |0 \rangle_B
= |0 \rangle_B, B''|1 \rangle_B = -|1 \rangle_B,\end{equation}for
the $B''$-representation $\{|0 \rangle_B,|1 \rangle_B \}$ $(0\leq
\beta \leq 2 \pi).$

We write $|00 \rangle_{AB},$ etc., as shorthand for $|0 \rangle_A
\otimes |0 \rangle_B.$ Since $\{ |00 \rangle_{AB}, |01
\rangle_{AB}, |1 0\rangle_{AB}, |11 \rangle_{AB} \}$ is a
orthogonal basis of the two-qubit system, we can uniquely write
$$| \psi \rangle = \lambda_{00}|0 0 \rangle_{AB} + \lambda_{01}
|0 1 \rangle_{AB} + \lambda_{10}|1 0 \rangle_{AB} + \lambda_{11}|1
1 \rangle_{AB},$$where$$|\lambda_{00}|^2 +
|\lambda_{01}|^2+|\lambda_{10}|^2 + |\lambda_{11}|^2 = 1.$$Since
$\psi$ maximize ${\cal B}_2,$ it also maximize ${\cal B}^2_2 =
4+4A''B''$ and so\begin{equation}A''B'' | \psi \rangle = | \psi
\rangle.\end{equation}By Eqs.(11) and (16) we have that
$\lambda_{01} =\lambda_{10}=0.$ Hence $| \psi \rangle$ is of the
form$$| \psi \rangle= a | 00 \rangle_{AB} + b | 11 \rangle_{AB}$$
with $|a|^2 + |b|^2 =1.$

On the other hand, we conclude by Eq.(6) that\begin{equation}\left
( AB + AB' + A'B - A'B' \right )| \psi \rangle = 2 \sqrt{2} | \psi
\rangle.\end{equation}By using Eqs.(12)-(15) we have that$$b
e^{i(\alpha + \beta)} (1-i)=\sqrt{2} a, ~a e^{-i(\alpha +
\beta)}(1+i) = \sqrt{2} b,$$and so$$ a =\frac{1}{\sqrt{2}} e^{i
(\alpha + \beta +\theta - \frac{\pi}{4})}, b = \frac{1}{\sqrt{2}}
e^{i \theta},$$where $0 \leq \theta \leq 2 \pi.$ Thus,
$$| \psi \rangle= e^{i \theta}\frac{1}{\sqrt{2}} \left ( e^{i (\alpha + \beta -
\frac{\pi}{4})}| 00 \rangle_{AB} +  | 11 \rangle_{AB} \right
).$$

Let $U_1$ be the unitary transform from the original
$\sigma^1_z$-representation to $A''$-representation on the first
qubit, i.e., $U_1 |0 \rangle = |0 \rangle_A$ and $U_1 |1 \rangle =
|1 \rangle_A,$ and similarly $U_2$ on the second qubit.
Define$$U_A = e^{i \theta} U_1 ,~~U_B =U_2 \left(
\begin{array}{ll}e^{ i(\alpha+\beta- \frac{\pi}{4})} & 0
\\~~~~ 0 & 1 \end{array} \right ).$$Then $U_A$ and
$U_B$ are unitary operators on the first qubit and the second
respectively, so that$$| \psi \rangle = \left ( U_A U_B \right )
\frac{1}{\sqrt{2}}\left ( | 00 \rangle + | 11 \rangle \right
),$$i.e., $| \psi \rangle$ can be obtained by a local unitary
transformation of the Bell state $\frac{1}{\sqrt{2}}\left ( | 00
\rangle + | 11 \rangle \right ).$

For the three-qubit system, let us consider a system of three
qubits labelled by $1, 2,$ and $3.$ Let $A,A'$ denote spin
observables on the first qubit, $B,B'$ on the second, and $C,C'$
on the third. Recall that the Bell-Klyshko inequality [11] for
three qubits reads that\begin{equation}\langle A' B' C + A' B C' +
A B' C' - A B C \rangle \leq 2,\end{equation}which holds true when
assuming EPR's local realism [2].

We define the three-qubit Bell operator [15]
\begin{equation}{\cal B}_3 = A' B' C+ A' B C' + A B' C' - A B
C.\end{equation}A simple computation yields that
\begin{widetext}\begin{equation}{\cal B}^2_3 =4 - [A,A'][B,B']-
[A,A'][C,C']- [B,B'][C,C']= 4 + 4[ (A \times A') (B \times B')+(A
\times A')(C \times C')+(B \times B')(C \times C')
].\end{equation}\end{widetext}Accordingly, by Eq.(4) we have
$\|{\cal B}^2_3 \| \leq 16$ and so $\|{\cal B}_3 \| \leq 4.$ As
follows, we will prove that a state $|\psi \rangle$ maximally
violates the Bell-Klyshko inequality Eq.(19),
i.e.,\begin{equation}\langle \psi | {\cal B}_3 | \psi \rangle =
4,\end{equation}if and only if it is of the form\begin{equation}|
\psi \rangle = \left ( U_A U_B U_C\right ) \frac{1}{\sqrt{2}}\left
( | 000 \rangle + | 111 \rangle \right )\end{equation}where $U_A,
U_B,$ and $U_C$ are unitary operators respectively on the first
qubit, second one and third one. Therefore, the GHZ state$$|GHZ
\rangle = \frac{1}{\sqrt{2}}\left ( | 000 \rangle + | 111 \rangle
\right )$$ and the states obtained from it by local unitary
transformations are the unique states that violate maximally the
Bell-Klyshko inequality of three qubit, as conjectured by Gisin
and Bechmann-Pasquinucci [8].

It is easy to check that all states of the form Eq.(23) maximally
violate Eq.(19). In this case, we only need to choose that$$A =
U_A \sigma^A_z \sigma^A_x \sigma^A_z U^*_A, A' = U_A \sigma^A_z
\sigma^A_y \sigma^A_z U^*_A;$$$$B = U_B \sigma^B_x U^*_B, B' = U_B
\sigma^B_y U^*_B;$$and$$C = U_C \sigma^C_x U^*_C, C' = U_C
\sigma^C_y U^*_C.$$This is so, because of that $|GHZ \rangle$
satisfies Eq.(22) for the Bell operator\begin{widetext}$${\cal
B}_3 = ( \sigma^A_z \sigma^A_y \sigma^A_z)\sigma^B_y \sigma^C_x+ (
\sigma^A_z \sigma^A_y \sigma^A_z) \sigma^B_x \sigma^C_y +
(\sigma^A_z \sigma^A_x \sigma^A_z) \sigma^B_y \sigma^C_y - (
\sigma^A_z \sigma^A_x \sigma^A_z) \sigma^B_x
\sigma^C_x.$$\end{widetext}Here $\vec{\sigma}_1 = \left (
\sigma^A_x, \sigma^A_y, \sigma^A_z \right ),\vec{\sigma}_2 = \left
( \sigma^B_x, \sigma^B_y, \sigma^B_z \right )$ and $\vec{\sigma}_3
= \left ( \sigma^C_x, \sigma^C_y, \sigma^C_z \right )$ are the
Pauli matrices for qubits $1,2$ and $3$ respectively.

The key point of the proof is that the converse holds true, i.e.,
the maximal violation of the Bell-Klyshko inequality of three
qubit occurs only for the GHZ state and the states obtained from
it by local unitary transformations, as similar as the two-qubit
case. We begin with the fact that if there exists a state $|\psi
\rangle$ satisfying\begin{equation}{\cal B}^2_3 |\psi \rangle = 16
|\psi
\rangle,\end{equation}then\begin{equation}(A,A')=(B,B')=(C,C')=0.
\end{equation}In fact, if Eq.(24) holds true, then $\|{\cal B}^2_3\|
= 16.$ By Eqs.(4) and (21) we immediately conclude Eq.(25). In the
sequel, we show that a state $|\psi \rangle$ satisfying Eq.(24)
must be of the form\begin{equation}|\psi \rangle = a | 0 \rangle_A
| 0 \rangle_B |0 \rangle_C + b | 1 \rangle_A |1 \rangle_B | 1
\rangle_C,\end{equation}$(|a|^2 + |b|^2 =1)$where $\{| 0
\rangle_A, | 1 \rangle_A \}, \{| 0 \rangle_B, | 1 \rangle_B \},$
and $\{| 0 \rangle_C, | 1 \rangle_C \}$ are orthogonal bases
respectively on the first qubit, second one, and third one.

Let $A''=A \times A',$ $B''=B \times B',$ and $C''=C \times C'.$
Since Eq.(25) holds true, as shown above, $\{A,A',A''\},$
$\{B,B',B''\},$ and $\{C,C',C''\}$ all satisfy the algebraic
identities Eqs.(7)-(10) that Pauli's matrices satisfy. By choosing
the $A''$-representation on the first qubit, $B''$-representation
on the second, and $C''$-representation on the third respectively,
we have Eqs.(11)-(16) and\begin{equation}C |0 \rangle_C =e^{-i
\gamma} |1 \rangle_C, C|1 \rangle_C = e^{i \gamma} |0
\rangle_C,\end{equation}
\begin{equation}C' |0
\rangle_C = ie^{-i \gamma} |1 \rangle_C, C' |1 \rangle_C =-i e^{i
\gamma} |0 \rangle_C,\end{equation}
\begin{equation}C'' |0 \rangle_C
= |0 \rangle_C, C''|1 \rangle_C = -|1 \rangle_C,\end{equation}for
the $C''$-representation $\{|0 \rangle_C,|1 \rangle_C \}$ $(0\leq
\gamma \leq 2 \pi).$

We write $|001 \rangle_{ABC},$ etc., as shorthand for $|0
\rangle_A \otimes |0 \rangle_B \otimes |1 \rangle_C.$ Since $\{
|\epsilon_A \epsilon_B \epsilon_C \rangle_{ABC}: \epsilon_A,
\epsilon_B, \epsilon_C = 0,1\}$ is a orthogonal basis of the
three-qubit system, we can uniquely write$$|\psi \rangle =
\sum_{\epsilon_A, \epsilon_B, \epsilon_C = 0,1}\lambda_{\epsilon_A
\epsilon_B \epsilon_C} |\epsilon_A \epsilon_B \epsilon_C
\rangle_{ABC}$$with $\sum \left |\lambda_{\epsilon_1 \epsilon_2
\epsilon_3} \right |^2 =1.$ By using Eqs.(11), (16) and (29), we
follow from Eq.(24) that
$$\lambda_{001}=\lambda_{010}=\lambda_{100}=\lambda_{011}=\lambda_{101}
=\lambda_{110}=0.$$Thus $|\psi \rangle = \lambda_{000} | 000
\rangle_{ABC} + \lambda_{111} | 111 \rangle_{ABC}$ is of the form
Eq.(26).

Now suppose that a state $|\psi \rangle$ satisfies Eq.(22). Since
Eq.(22) is equivalent to that\begin{equation} {\cal B}_3 | \psi
\rangle = 4 | \psi \rangle,\end{equation}it concludes that $|\psi
\rangle$ satisfies Eq.(24) and hence is of the form Eq.(26). Note
that$${\cal B}_3 = {\cal B}_2 \otimes \frac{1}{2} \left( C + C'
\right ) + {\cal B}'_2 \otimes \frac{1}{2} \left( C - C' \right
)$$where ${\cal B}'_2 = A'B' + A'B + AB' - AB$ denote the same
expression ${\cal B}_2$ but with the $A$ and $A',$ $B$ and $B'$
exchanged. Then, by Eqs.(27) and (28) one has
that\begin{widetext}$${\cal B}_3 | \psi \rangle = \frac{1}{2}a
e^{-i \gamma} \left [ (1+i) {\cal B}_2 + (1-i) {\cal B}'_2 \right
] |00 \rangle_{AB} |1 \rangle_C + \frac{1}{2}b e^{ i \gamma} \left
[ (1-i) {\cal B}_2 + (1+i) {\cal B}'_2 \right ] |11 \rangle_{AB}
|0 \rangle_C.$$\end{widetext}From Eq.(30) we conclude
that\begin{equation}\frac{1}{2}a e^{-i \gamma}\left [ (1+i) {\cal
B}_2 + (1-i) {\cal B}'_2 \right ]|0 0\rangle_{AB} =
4b|11\rangle_{AB},\end{equation}and
\begin{equation}\frac{1}{2}b
e^{i \gamma} \left [ (1-i) {\cal B}_2 + (1+i) {\cal B}'_2 \right
]|11\rangle_{AB} = 4a|00\rangle_{AB}.\end{equation}By Eq.(31) we
have that$$\begin{array}{lcl}4|b| & \leq & \frac{1}{2}|a| \left (
|1+i| \| {\cal B}_2 \| + |1-i| \|{\cal B}'_2 \| \right )\\& \leq &
\frac{1}{2}|a|\left ( \sqrt{2}\cdot 2\sqrt{2} + \sqrt{2}\cdot
 2\sqrt{2} \right )\\&=& 4 |a|,\end{array}$$since $\|
{\cal B}_2 \|, \| {\cal B}'_2 \|\leq 2\sqrt{2}$ as shown in the
two-qubit case. This concludes that $|b| \leq |a|.$ Similarly, by
Eq.(32) we have that $|a| \leq |b|$ and so $|a|=|b|.$ Therefore,
we have that $a=\frac{1}{\sqrt{2}} e^{i \phi}, b =
\frac{1}{\sqrt{2}} e^{i \theta}$ for some $0 \leq \phi, \theta
\leq 2 \pi,$ that is,$$| \psi \rangle = \frac{1}{\sqrt{2}} \left (
e^{i \phi}| 000 \rangle_{ABC} + e^{i \theta} | 111 \rangle_{ABC}
\right ).$$This immediately concludes that $| \psi \rangle$ can be
obtained by a local unitary transformation of the GHZ state,
precisely$$| \psi \rangle = \left ( U_A U_B U_C\right )
\frac{1}{\sqrt{2}}\left ( | 000 \rangle + | 111 \rangle \right
),$$where$$U_A = U_1 \left (\begin{array}{ll} e^{ i\phi} & 0
\\~ 0 & 1 \end{array} \right ),~~U_B = U_2 \left(
\begin{array}{ll}1 & ~0
\\ 0 & e^{i \theta} \end{array} \right )$$ and $U_C = U_3.$
Here, $U_1$ is the unitary transform from the original
$\sigma^1_z$-representation to $A''$-representation on the first
qubit, i.e., $U_1 |0 \rangle = |0 \rangle_A$ and $U_1 |1 \rangle =
|1 \rangle_A,$ and similarly, $U_2$ on the second qubit and $U_3$
on the third qubit, respectively.

To sum up, by using some subtle mathematical techniques we have
shown that the Bell and GHZ states and the states obtained from
them by local unitary transformations are the unique states that
violate maximally the Bell-CHSH and Bell-Klyshko inequalities,
respectively. This was conjectured by Gisin and
Bechmann-Pasquinucci [8]. The key point of our argument involved
here is by using the certain algebraic properties that Pauli's
matrices satisfy, which is based on the determination of local
spin observables of the associated Bell operator. The method
involved here is simpler (and more powerful) than one used in [9]
and can be extended to the $n$-qubit case, which will be presented
elsewhere. It is known that maximally entangled states should
maximally violate the Bell inequality and all states obtained by
local unitary transformations of a maximally entangled state are
equally valid entangled states [12], we therefore obtain the
characterizations of maximally entangled states in both two-qubit
and three-qubit via the Bell-CHSH and Bell-Klyshko inequalities,
respectively. Finally, we remark that the W state of
three-qubit$$|W \rangle = \frac{1}{3}\left (|001 \rangle + |010
\rangle + |100 \rangle \right )$$cannot be obtained from the GHZ
state by a local unitary transformation and hence does not
maximally violate the Bell-Klyshko inequality, although it is a
``maximally entangled" state in the sense described in [16]. This
also occurs in the GHZ theorem [5], that is, the W state does not
provide an $100 \%$ contradiction between quantum mechanics and
EPR's local realism [17]. Since the Bell inequality and GHZ
theorem are two main theme on the violation of EPR's local
realism, it concludes that from the point of view on the violation
of EPR's local realism, the W state cannot be regarded as a
``maximally entangled" state and hence we need some new ideas for
clarity of the W state [18].


\end{document}